\documentclass[usenatbib]{mn2e}
\usepackage{upgreek}
\usepackage{amssymb,amsmath}
\usepackage{graphicx}
\usepackage{multirow}
\usepackage{natbib}
\usepackage{tabularx}
\def\cite{\citep}

\usepackage[usenames,dvipsnames]{color}

\title[A real-time FRB]{A real-time fast radio burst: polarization detection and multiwavelength follow-up}
\makeatletter
\author[Petroff et al.]{E.~Petroff$^{1,2,3}$\thanks{Email: epetroff@astro.swin.edu.au}, M.~Bailes$^{1,3}$, E.~D.~Barr$^{1,3}$, B.~R.~Barsdell$^{4}$, N.~D.~R.~Bhat$^{3,5}$, F.~Bian$^{6,7}$, \newauthor S.~Burke-Spolaor$^8$, M.~Caleb$^{7,1,3}$, D.~Champion$^9$, P.~Chandra$^{10}$, G.~Da Costa$^7$, C.~Delvaux$^{11}$, \newauthor C.~Flynn$^{1,3}$, N.~Gehrels$^{12}$, J.~Greiner$^{11}$, A.~Jameson$^{1,3}$, S.~Johnston$^2$, M.~M.~Kasliwal$^{13,14}$, \newauthor  E.~F.~Keane$^{1,3}$, S.~Keller$^7$, J.~Kocz$^{4,15}$, M.~Kramer$^{9,16}$, G.~Leloudas$^{17,18}$, D.~Malesani$^{17}$, \newauthor J.~S.~Mulchaey$^{13}$, C.~Ng$^9$, E.~O.~Ofek$^{18}$, D.~A.~Perley$^{8,14}$, A.~Possenti$^{19}$, B.~P.~Schmidt$^{7,3}$,\newauthor Yue~Shen$^{13,20}$, B.~Stappers$^{16}$, P.~Tisserand$^{7,3,21,22}$, W.~van Straten$^{1,3}$, C.~Wolf$^{7,3}$ 
\\
$^1$Centre for Astrophysics and Supercomputing, Swinburne University of Technology, P.O. Box 218, Hawthorn, VIC 3122, Australia \\
$^2$CSIRO Astronomy \& Space Science, Australia Telescope National Facility, P.O. Box 76, Epping, NSW 1710, Australia \\
$^3$ARC Centre of Excellence for All-sky Astrophysics (CAASTRO) \\
$^4$Harvard-Smithsonian Center for Astrophysics, 60 Garden Street, Cambridge, Massachusetts, 02138, USA\\
$^5$International Centre for Radio Astronomy Research, Curtin University, Bentley, WA 6102, Australia \\
$^6$Stromlo Fellow \\
$^7$Research School of Astronomy and Astrophysics, Australian National University, ACT, 2611, Australia \\
$^8$Cahill Center for Astrophysics, California Institute of Technology, 1200 E California Blvd, Pasadena, CA 91125, USA\\
$^9$Max Planck Institut f\"{u}r Radioastronomie, Auf dem H\"{u}gel 69, D-53121 Bonn, Germany \\
$^{10}$National Centre for Radio Astrophysics, Tata Institute of Fundamental Research, Pune University Campus, Ganeshkhind, Pune 411 007, India \\
$^{11}$Max-Planck-Institut f\"{u}r extraterrestrische Physik, Giessenbachstrasse 1, 85748 Garching, Germany \\
$^{12}$Astrophysics Science Division, NASA Goddard Space Flight Center, USA \\
$^{13}$Observatories of the Carnegie Institution for Science, 813 Santa Barbara Street, Pasadena, CA 91101, USA \\
$^{14}$Hubble Fellow \\
$^{15}$Jet Propulsion Laboratory, California Institute of Technology, 4800 Oak Grove Drive, Pasadena, CA 91104, USA \\
$^{16}$Jodrell Bank Centre for Astrophysics, University of Manchester, Alan Turing Building, Oxford Road, Manchester M13 9PL, United Kingdom \\
$^{17}$Dark Cosmology Centre (DARK), Niels Bohr Institute, University of Copenhagen, Juliane Maries Vej 30, 2100 Copehagen \O, Denmark \\
$^{18}$Department of Particle Physics \& Astrophysics, Weizmann Institute of Science, Rehovot 76100, Israel \\
$^{19}$INAF - Osservatorio Astronomico di Cagliari, Via della Scienza 5, 09047 Selargius (CA), Italy \\
$^{20}$Kavli Institute for Astronomy and Astrophysics, Peking University, Beijing 100871, China \\
$^{21}$Sorbonne Universit\'{e}s, UPMC Univ Paris 06, UMR 7095, Institut d'Astrophysique de Paris, F-75005 Paris, France \\
$^{22}$CNRS, UMR 7095, Institut d'Astrophysique de Paris, 98 bis Boulevard Arago, F-75014 Paris, France
}
\makeatother

\let\leqslant=\leq %Authors with AMS fonts and mssymb.tex can comment
\date{}
\begin{document}

\maketitle

\begin{abstract}
Fast radio bursts (FRBs) are one of the most tantalizing mysteries of the radio sky; their progenitors and origins remain unknown and until now no rapid multiwavelength follow-up of an FRB has been possible. New instrumentation has decreased the time between observation and discovery from years to seconds, and enables polarimetry to be performed on FRBs for the first time. We have discovered an FRB (FRB 140514) in real-time on 14 May, 2014 at 17:14:11.06 UTC at the Parkes radio telescope and triggered follow-up at other wavelengths within hours of the event. FRB 140514 was found with a dispersion measure (DM) of 562.7(6) cm$^{-3}$ pc, giving an upper limit on source redshift of $z \lesssim 0.5$. FRB 140514 was found to be 21$\pm$7\% (3-$\sigma$) circularly polarized on the leading edge with a 1-$\sigma$ upper limit on linear polarization $<10\%$. We conclude that this polarization is intrinsic to the FRB. If there was any intrinsic linear polarization, as might be expected from coherent emission, then it may have been depolarized by Faraday rotation caused by passing through strong magnetic fields and/or high density environments. FRB 140514 was discovered during a campaign to re-observe known FRB fields, and lies close to a previous discovery, FRB 110220; based on the difference in DMs of these bursts and time-on-sky arguments, we attribute the proximity to sampling bias and conclude that they are distinct objects. Follow-up conducted by 12 telescopes observing from X-ray to radio wavelengths was unable to identify a variable multiwavelength counterpart, allowing us to rule out models in which FRBs originate from nearby ($z < 0.3$) supernovae and long duration gamma-ray bursts.  
\end{abstract}

\begin{keywords}
polarization --- radiation mechanisms: general --- intergalactic medium ---radio continuum: general
\end{keywords}

\section{Introduction}

A new class of objects called fast radio bursts (FRBs) have been discovered in radio pulsar surveys at Parkes and Arecibo within the last decade \cite{Lorimer07,Thornton13,Spitler14,SarahFRB}. All FRBs discovered to date have been single radio events of millisecond duration. The electron column density, called the dispersion measure (DM), is also uncharacteristically high, leading to theories that they originate at cosmological distances \cite{Thornton13} and/or in extreme environments \cite{Katz2014,Lyubarsky2014}. Recently, they have been the topic of considerable discussion, both as to their origins and their potential use as cosmological tools \cite{Loeb14,Kulkarni14,Deng2014,Gao14}.

\citet{Thornton13} measure a rate of R$_\mathrm{FRB}$($\mathcal{F} \sim$ 3 Jy ms) $\sim$ 1.0$\substack{+0.6 \\ -0.5} \times 10^{4}$ sky$^{-1}$ day$^{-1}$ from 4 events found in a high Galactic latitude search, but the non-detection of FRBs in a survey twice as long suggests either a lower overall FRB rate or a latitude dependence, owing to a currently unknown obscuration effect at Galactic latitudes below $|b| = 15^{\circ}$ \cite{Petroff14}. This result has recently been confirmed by \citet{SarahFRB}. The true progenitors of FRBs remain unknown. All published FRBs were discovered in archival data years later and rapid follow-up of an FRB has never been possible. 

Recent efforts in time-domain radio astronomy have focused on real-time FRB detection with the promise of rapid follow-up of new events. Such capability was recently made possible with the development of a real-time transient pipeline at the Parkes telescope. A new survey at Parkes aims to search the fields of previous FRB events for repeating bursts, a direct prediction made by flare star and magnetar flare origin theories \cite{Loeb14,Kulkarni14}. The discovery of repeating FRB sources would strongly constrain emission mechanisms and possible progenitors. 

Here we report on early results from this survey with the discovery of a new FRB in the field of FRB 110220. In $\S$\ref{sec:pipeline} we describe the real-time transient pipeline at Parkes using the multibeam receiver. In $\S$\ref{sec:radio} we present the detection of  FRB 140514 and, for the first time, the polarized radiation of an FRB. $\S$\ref{sec:followup} details the follow-up efforts from X-ray to radio from 12 observatories. We summarize the results from these follow-ups in $\S$\ref{sec:discussion} and discuss polarization in $\S$\ref{sec:polarization}, connections between FRB 140514 and FRB 110220 in $\S$\ref{sec: FRB 110220}, and limits on an afterglow in $\S$\ref{sec:afterglow}. We provide a conclusion in $\S$\ref{sec:conclusion}.

\section{Real-Time Transient Pipeline}\label{sec:pipeline}

Observations were conducted with the Berkeley Parkes Swinburne Recorder (BPSR) backend for the 13-beam multibeam receiver \cite{multibeam} at Parkes which covers 0.5 deg$^2$ on the sky. We record 8-bit full-polarization data from two orthogonal linear feeds per beam, with 1024 frequency channels over 400 MHz of bandwidth, from $1182 - 1582$ MHz, and 64-$\upmu$s time resolution. Data are passed to the HI-Pulsar Signal Processor (HIPSR) where 120 s of observations are stored in a ring buffer using PSRDada\footnote{http://psrdada.sourceforge.net}. The effective bandwidth for our data is 340 MHz from 1182 to 1522 MHz due to communications satellites operating in the 1525 to 1559 MHz band \cite{Keith10}. The observing instrumentation is identical to that used for the High Time Resolution Universe (HTRU) survey and the FRB discoveries reported in \citet{Thornton13}.

The real-time processing of the data for transient events is performed on the buffer using the \textsc{Heimdall} single pulse processing software\footnote{http://sourceforge.net/projects/heimdall-astro/}. The linear polarizations are summed into a single 8-bit data set and are passed to \textsc{Heimdall} in 256 kilosample chunks (approximately 16.77 s). If a candidate is detected, the relevant 8-bit data are saved to disk. \textsc{Heimdall} performs a search for pulses across a specified range of DMs and pulse widths and returns a list of candidates. The real-time transient pipeline searches $0-2000$ cm$^{-3}$ pc in DM and $0.128-262$ ms in pulse width and identifies candidates that fit a number of criteria attributable to known FRBs:

\begin{equation}\label{eq:frblimits}
\begin{split}
\mathrm{DM} \geq 1.5 \times \mathrm{DM}_\mathrm{MW} \\
\mathrm{S/N} \geq 10 \\
N_\mathrm{beams} \leq 4 \\
\Delta t \leq 8.192 \: \mathrm{ms} \\
N_\mathrm{events}(t_\mathrm{obs}-2\:\mathrm{s} \to t_\mathrm{obs}+2\:\mathrm{s}) \leq 5
\end{split}
\end{equation}
\noindent where DM$_\mathrm{MW}$ is the Galactic DM along the line of sight predicted by the NE2001 electron density model \cite{Cordes02}, S/N is the signal to noise ratio, $N_\mathrm{beams}$ is the number of beams in which the candidate appears, $\Delta t$ is the pulse width, and $N_\mathrm{events}(t_\mathrm{obs}-2\:\mathrm{s} \to t_\mathrm{obs}+2\:\mathrm{s})$ is the number of separately identified pulse candidates within a 4 s window around the candidate, more than the expected time duration of a single burst with DM $\leq$ 2000 pc cm$^{-3}$. For each candidate which meets all criteria above, we also check that there is no known pulsar in the pulsar catalogue \cite{psrcat} within a 5\% DM range of the candidate for completeness, although this condition is typically precluded by the high-DM threshold. All known archival FRBs are identified using these criteria.

When a candidate is detected the observable time span of the event is calculated using the total DM delay, $\uptau_\mathrm{DM}$, across our observing bandwidth:

\begin{equation}\label{eq:DMsmear}
\uptau_\mathrm{DM} = 4.15 \: \mathrm{DM} \left( \nu_\mathrm{low}^{-2} - \nu_\mathrm{high}^{-2} \right) \: \mathrm{ms}
\end{equation}
\noindent where $\nu_\mathrm{low}$ and $\nu_\mathrm{high}$ are the lowest and highest frequencies in the band, in GHz, and DM is in pc cm${^-3}$. For all BPSR data $\nu_\mathrm{low}$ = 1.182 GHz and $\nu_\mathrm{high}$ = 1.582 GHz. The start time of the event is identified in the 120 s buffer and all samples in the range ($t_\mathrm{start}$-$\uptau_\mathrm{DM}$, $t_\mathrm{start}$+2$\uptau_\mathrm{DM}$) are saved to disk with full polarization 8-bit data. 

The BPSR real-time candidate detection and polarization triggering mode was commissioned in March, 2014. Previously, it was impossible to obtain polarization data for FRBs at Parkes. Currently the triggers are configured to give a few false positives rather than miss a real event, and the real-time nature of the pipeline enables the trained observer to provide immediate feedback. The triggers have not yet been connected directly with other telescopes, but instead only initiate an email alert to observers related to the project when an event satisfying the above criteria is found.

\section{Parkes real-time detection of FRB 140514}\label{sec:radio}

FRB 140514 was discovered on 14 May, 2014 at 17:14:11.06 UTC (15 May 03:14:11.06 local time) at 1.4~GHz in the centre beam (beam 01) of the multibeam receiver. It was identified in the \textsc{Heimdall} real-time transient pipeline with an S/N of 16, a DM of 562.7(6) pc cm$^{-3}$, and a pulse width of 2.8$\substack{+3.5 \\ -0.7}$ ms.  The pipeline identified the burst within 10 seconds and 2.22 seconds of data around the event were recorded to disk in 8-bit dual polarization for all 13 beams of the receiver. An FRB alert email was sent to project observers at 17:14:30 UTC.

If the burst occurred at the beam-centre the detection corresponds to a peak flux density of 0.47 $^{+0.11}_{-0.08}$ Jy and a fluence of 1.3 $^{+2.3}_{-0.5}$ Jy ms. Further analysis of the FRB data resulted in a dispersion index $\alpha = -2.000(4)$ such that $\delta t \propto \mathrm{DM} \, \nu^\alpha$, in agreement with the $\nu^{-2}$ expected for cold plasma. The scattering time-scale was found to be $\tau_\mathrm{1GHz}$ = 5.4(1) ms. There is a decreased uncertainty in the dispersion index and the scattering time-scale, as the scattering index was not a free parameter in the fit algorithm. While the scattering modelling done for these fits is consistent with a range of different pulse widths, leading to large error in $\Delta t$, the effect on the scattering tail, and thus $\tau_\mathrm{1GHz}$, is negligible, giving a smaller error.

\begin{figure}
\includegraphics[width=8cm]{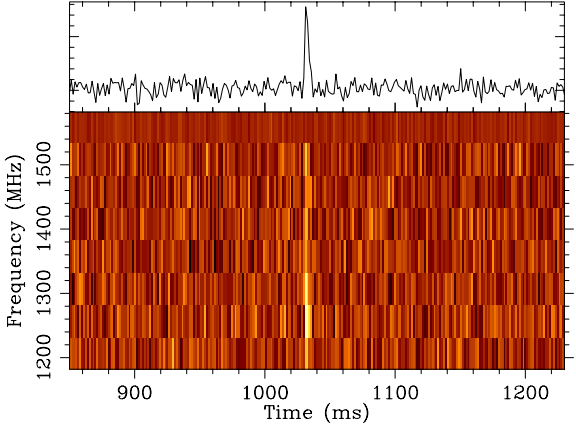}
\caption{The pulse profile and dynamic spectrum of FRB 140514 with pulse width 2.8$^{+3.5}_{-0.7}$ ms, dedispersed to DM = 562.7 pc cm$^{-3}$ and summed to 8 frequency channels across the band. The total time plotted has been reduced to 400 ms for greater clarity. Frequency channels between 1520 to 1580 MHz are excised due to narrow-band radio interference from the Thuraya 3 satellite which transmits in this band.}\label{fig:waterfall}
\end{figure}

The DM, dispersion index and scattering time-scale were all fit for while the scattering index was held fixed at $\beta$ = -4. The limited S/N (16) of the pulse prohibited fitting for $\beta$ due to strong covariances between the four quantities. See Figure~\ref{fig:waterfall} and Table~\ref{tab:frb} for all observed FRB parameters, and Table~\ref{tab:cosmology} for derived cosmological parameters. 

All 13 beams of full-Stokes data were analysed in detail and the pulse was not detected in any other beam of the receiver. Since there was no coincident detection in other beams, we conclude that the event was not a sidelobe detection. Therefore we have used the coordinates from the beam center for the detection pointing with an error diameter of 14.4$'$, the approximate full-width half-maximum (FWHM) of beam 01 at 1.4~GHz \cite{multibeam}.

FRB 140514 was discovered in a pointing centered just 9$'$ away from the nominal position of known FRB 110220 during a standard gridding \cite{gridding} of the region in our survey. The previous event nearby, FRB 110220, had DM = 944.38(5) pc cm$^{-3}$ and a peak flux density of 1.3 Jy, if it occurred at beam-center.

Models of the free electron content of the Milky Way predict that the ionized Galactic interstellar medium contribution to the DM of FRB 140514 is only 35 pc cm$^{-3}$ \cite{Cordes02}, only 6\% of the total, which sets an upper limit on redshift $z < 0.4(1)$ based on ionization models of the intergalactic medium (IGM), making no assumptions about a host contribution to the total DM, and assuming an upper limit on the Galactic DM contribution of 70 pc cm$^{-3}$ \cite{Ioka03}. This upper limit on redshift corresponds to a co-moving distance of $<$ 1.71(3) Gpc, a luminosity distance of $<$ 2.46$^{+0.04}_{-0.06}$ Gpc, an energy of $<$ 3.7$^{+4.7}_{-2.0} \times 10^{38}$ erg, and a distance modulus of $<$ 42.2 mag \cite{CosmologyCalc}. In comparison, the upper limit on redshift for FRB 110220 was $z < 0.81$, which corresponds to a co-moving distance of 2.8 Gpc, a luminosity distance of 5.1 Gpc, and a distance modulus of 43.5 mag \cite{Thornton13} if most of the excess DM is attributed to the IGM.

\begin{table}
\setlength\extrarowheight{3pt}
\begin{centering}
\caption{Observed properties of FRB 140514}\label{tab:frb}
\begin{tabular}{cc}
\hline
\hline
Event date UTC & 14 May, 2014 \\
Event time UTC, $\nu_\mathrm{1.4~GHz}$ & 17:14:11.06 \\
Event time, $\nu_\infty$ & 17:14:09.83 \\
Local date AEST & 15 May, 2014 \\
Local time AEST & 03:14:11.06 \\
RA & 22:34:06.2 \\
Dec & $-$12:18:46.5\\
($\ell$,$b$) & (50.8$^{\circ}$, $-$54.6$^{\circ}$)\\
Beam diameter & 14.4$'$ \\
DM$_\mathrm{FRB}$ (pc cm$^{-3}$) & 562.7(6) \\
DM$_\mathrm{MW}$ (pc cm$^{-3}$) & 34.9 \\
Detection S/N & 16(1) \\ %Errors updated to agree with E&E
Observed width, $\Delta t$ (ms) & 2.8 $\substack{+3.5 \\ -0.7}$ \\ %% Errors updated to agree with E&E
Scattering timescale, $\tau_\mathrm{1GHz}$ (ms) & 5.4(1) \\
Dispersion index, $\alpha$ & -2.000(4) \\
Peak flux density, $S_{\nu,\mathrm{1400MHz}}$ (Jy) & 0.47 $\substack{+0.11 \\ -0.08}$ \\ %% Errors updated to agree with E&E
Fluence, $\mathcal{F}$ (Jy ms) & 1.3 $\substack{+2.3 \\ -0.5}$ \\
\hline
\end{tabular}
\end{centering}
\end{table}

\begin{table}
\setlength\extrarowheight{3pt}
\begin{centering}
\caption{Derived cosmological properties of FRB 140514}\label{tab:cosmology}
\begin{tabular}{cc}
\hline
\hline
$z$ & $<$ 0.44(1) \\
Co-moving distance (Gpc) & $<$ 1.71(3) \\
Luminosity distance (Gpc) & $<$ 2.46$^{+0.04}_{-0.06}$ \\
Energy (erg) & $<$ 3.7$^{+4.7}_{-2.0} \times 10^{38}$ \\
Distance modulus (mag) & $<$ 42.2 \\
\hline
\end{tabular}
\end{centering}
\end{table}

\begin{figure*}
\includegraphics[width=16cm]{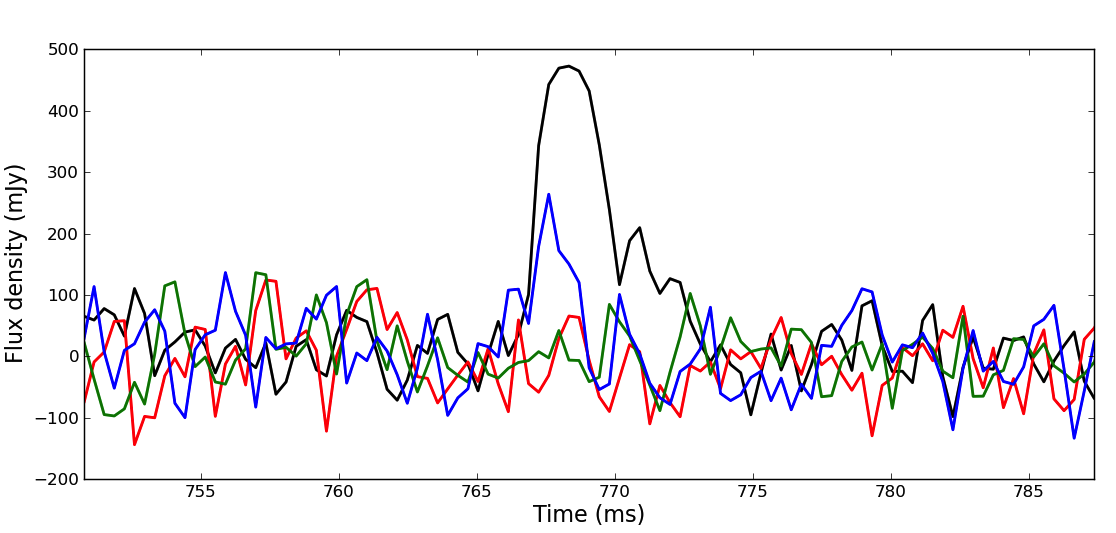}
\caption{The full-Stokes parameters of FRB 140514 recorded in the centre beam of the multibeam receiver with BPSR. Total intensity, and Stokes $Q$, $U$, and $V$ are represented in black, red, green, and blue, respectively. FRB 140514 has 21 $\pm$ 7\% (3-$\sigma$) circular polarisation averaged over the pulse, and a 1-$\sigma$ upper limit on linear polarisation of $L$ $<$ 10\%. On the leading edge of the pulse the circular polarisation is 42 $\pm$ 9\% (5-$\sigma$) of the total intensity. The data have been smoothed from an initial sampling of 64 $\upmu$s using a Gaussian filter of full-width half-maximum 90 $\upmu$s.}\label{fig:pol}
\end{figure*}

A calibration observation was taken at the end of the observing session at 01:04:39 UTC on 15 May, 7h50m after FRB 140514, which was used to calibrate the polarized data. The feed was assumed to be ideal and the calibration was performed using the \texttt{pac} command in the \textsc{PSRCHIVE} software package\footnote{http://psrchive.sourceforge.net/index.shtml}  \cite{Hotan2004}. We did not perform a Mueller matrix calculation as we cannot determine the exact location of the FRB within the Parkes beam. From the calibration of the orthogonal linear feeds we obtained all four Stokes parameters, plotted in Figure~\ref{fig:pol}; this represents the first detection of polarized flux from an FRB.

The emission of FRB 140514 was polarised with 21 $\pm$ 7\% (3-$\sigma$) circular polarisation averaged over the whole pulse. On the leading edge of the pulse, however, the pulse is 42 $\pm$ 9\% circularly polarised, a 5-$\sigma$ detection. No linear polarisation was detected and we place a 1-$\sigma$ upper limit of 10\% of the total intensity (Figure~\ref{fig:pol}). We note that it would require a very rare and specific feed rotation to result in high fractional circular polarisation with no linear detection. Such a configuration would also result in a high correlation between Stokes $V$ and $I$, and we do not observe the circular polarisation to tightly follow the total intensity. The measured circular polarisation is determined to be intrinsic to the observation, and not a calibration artefact.

With a polarized signal it is possible to measure the Faraday rotation of the Stokes vectors as a function of frequency due to the magnetic field and electron column density along the line of sight. The amount of induced rotation is quantified by the rotation measure,

\begin{equation}\label{eq:rm}
RM \propto \int_0^d n_e B_\parallel \mathrm{d}l \:,
\end{equation}

\noindent where $d$ is the distance to the source, $n_e$ is the electron column density, and $B_\parallel$ is the magnetic field parallel to the line of sight such that the rotation angle of the linear polarization $\Delta \Psi = \mathrm{RM} \lambda^2$ \cite{PulsarHandbook}. An optimal rotation measure search was performed using the \texttt{rmfit} code in the \textsc{psrchive} pulsar software package out to $|\mathrm{RM}_\mathrm{max}| = 1.18 \times 10^{5}$ rad m$^{-2}$, the RM at which the signal is completely depolarized within a single frequency channel at our observing frequency. No linear polarization was evident at $\geq 3\sigma$ significance.

\section{FRB Follow-up at Other Telescopes}\label{sec:followup}

The real-time detection of FRB 140514 enabled an extensive coordination of telescopes and multiwavelength observations. Three sources of interest were identified in the 14.4$'$ diameter of the Parkes beam: two X-ray sources detected by \textit{Swift}, referred to below as XRT1 and XRT2, and one from the Giant Metrewave Radio Telescope (GMRT), referred to as GMRT1. The search for a counterpart focused on identification of any variable slow transients in the field that either brightened or dimmed by $>$ 2 mag between epochs. For objects near the magnitude limit of the observation (the magnitude of the dimmest detectable source in the observation) an appearance or disappearance between epochs was only considered significant if the source was 2 magnitudes or more above the limit in one observing epoch. Ultimately, no afterglow-like counterpart was identified at any wavelength involved in this effort. Here we report on the findings from twelve telescopes involved in the follow-up effort, listed in Table~\ref{tab:followup}.

\subsection{Parkes Radio Telescope}

FRB 140514 was discovered in the first grid pointing around the position of FRB 110220 of the observing session, 9$'$ away from the previous FRB discovery position, less than one beamwidth \cite{Thornton13}. A grid of the field was observed for another 1.6 hours after detection as part of the scheduled observations, and then again at the end of the observing session, 7 hours after discovery, both with 145 mJy rms at 1.4~GHz. No further dispersed pulses at DM $\geq$ 5 pc cm$^{-3}$ were detected in subsequent observations. No other FRB-like pulses were found throughout the observing session along any other sightlines, and there was no strong radio frequency interference (RFI). The field was re-observed for 8 consecutive hours on 24 June, 41 days after the FRB event and again for 3.5 hours on 27 July, 74 days after FRB 140514, and no new candidates were identified.

\subsection{Australia Telescope Compact Array}

The Australia Telescope Compact Array (ATCA) observed the FRB field as a target of opportunity (ToO, proposal CX293) starting at 00:10 UT on 2014 May 15, less than 7 hours after the FRB. The total observing time was 3 hours including calibration and overheads. Observations were made simultaneously from 4.5-6.5 GHz and 8-10 GHz for a total of 60 min on source and from 1.1-3.1 GHz with 60 min on source. The rms of the images are approx 40 $\mu$Jy at the higher frequencies and 60 $\mu$Jy at 2 GHz. GMRT1 was identified in the ATCA image with a flux density of 1.5 mJy and XRT1 was also visible with a flux density of 3 mJy. While the ATCA was the first telescope other than Parkes to image the field, the lack of a second epoch days or weeks later hampered our ability to detect variable sources at these radio frequencies. No radio source could be targeted based on the ATCA observations.

\subsection{Giant Metrewave Radio Telescope}

The Giant Metrewave Radio Telescope (GMRT) began observing the FRB field as a target of opportunity (ToO, proposal DDT B124) 2 days after the event at 610~MHz at 01:30 UT on 16 May. The 3-hour observation (including overheads and calibrators) produced an image of the field with 123 $\upmu$Jy rms. We identified three sources within the field of view (J2000 coordinates): GMRT1 (RA = 22:34:08.493, Dec = $-$12:18:27.00), GMRT2 (RA = 22:34:19.003, Dec = $-$12:21:30.38), and GMRT3 (RA = 22:34:00.088, Dec = $-$12:14:50.00). The first, GMRT1, did not appear to correspond to any sources in the NRAO VLA Sky Survey catalog (NVSS, \citealt{NVSS}), and was flagged for further follow-up by other telescopes as a potentially variable source given the temporal proximity of the GMRT observation and the FRB detection. The other two sources, GMRT2 and GMRT3, correlated well with positions for known radio sources in the NVSS catalog with consistent flux densities. Subsequent observations were taken through the GMRT ToO queue on 20 May, 3 June, and 8 June in the 325~MHz, 1390~MHz, and 610~MHz bands, respectively. The second epoch was largely unusable due to technical difficulties. The search for variablility focused on monitoring each source for flux variations across observing epochs. All sources from the first epoch appeared in the third and fourth epochs with no measureable change in flux densities. 

%revisit Sarah's comments here

\subsection{\textit{Swift} X-Ray Telescope}

The first observation of the FRB 140514 field was taken using \textit{Swift} XRT \cite{swift} only 8.5 hours after the FRB was discovered at Parkes. This was the fastest \textit{Swift} follow-up ever undertaken for an FRB. 4 ks of XRT data were taken in the first epoch, and a further 2 ks of data were taken in a second epoch later that day, 23 hours after  FRB 140514, to search for short term variability. A final epoch, 18 days later, was taken to search for long term variability. Two X-ray sources were identified in the first epoch of data within the 15$'$ diameter of the Parkes beam. Both sources were consistent with sources in the USNO catalog \cite{USNOcatalog}. The first source (XRT1) is located at RA = 22:34:41.49, Dec = -12:21:39.8 with $R_\mathrm{USNO}$ = 17.5 and the second (XRT2) is located at RA = 22:34:02.33 Dec = -12:08:48.2 with $R_\mathrm{USNO}$ = 19.7. Both XRT1 and XRT2 appeared in all subsequent epochs with no observable variability on the level of 10\% and 20\% for XRT1 and XRT2, respectively, both calculated from photon counts from the XRT. Both sources were later found to be active galactic nuclei (AGN). 

\subsection{Gamma-Ray Burst Optical/Near-Infrared Detector} 

After 13 hours, a trigger was sent to the Gamma-Ray Burst Optical/Near-Infrared Detector (GROND) operating on the 2.2-m MPI/ESO telescope on La Silla in Chile \cite{grond}. GROND is able to observe simultaneously in $J$, $H$, and $K$ near-infrared (NIR) bands with a $10'\times10'$ field of view (FOV) and the optical $g'$, $r'$, $i'$, and $z'$ bands with a $6'\times6'$ FOV. A 2$\times$2 tiling observation was done, providing 61\% ($JHK$) and 22\% ($g'r'i'z'$) coverage of the inner part of the FRB error circle. The first epoch began 16 hours after FRB 140514 with 460 second exposures, and a second epoch was taken 2.5 days after the FRB with an identical observing setup and 690 s ($g'r'i'z'$) and 720 s ($JHK$) exposures, respectively. Limiting magnitudes for $J$, $H$, and $K$ bands were 21.1, 20.4, and 18.4 in the first epoch and 21.1, 20.5, and 18.6 in the second epoch, respectively (all in the AB system). Of all the objects in the field, analysis identified three variable objects, all very close to the limiting magnitude and varying on scales of 0.2 - 0.8 mag in the NIR bands identified with difference imaging. Of the three objects one is a galaxy, another is likely to be an AGN, and the last is a main sequence star. Both XRT1 and GMRT1 sources were also detected in the GROND infrared imaging but were not observed to vary in the infrared bands on the timescales probed.

\subsection{Swope Telescope}

An optical image of the FRB field was taken 16h51m after the burst event with the 1-m Swope Telescope at Las Campanas. The field was re-imaged with the Swope Telescope on 17 May, 2 days after the FRB. No variable optical sources were identified in the observations field of view, 1.92 deg$^2$, to a limiting magnitude of $R=$ 16.

\subsection{Palomar Transient Factory}

The intermediate Palomar Transient Factory (iPTF) uses the 1.2-m Samuel Oschin Telescope at Palomar Observatory at $R$-band with 60 s exposure times to search for optical transients over 8.1 deg$^2$ \cite{ptf,Rau09}. The iPTF was triggered within 12 hours and observations began approximately 18 hours after FRB 140514 on the night of 15 May. Four epochs of suitable data were taken of the field on May 15, 16, 17, and 19 with an $R$-band limiting magnitude of 19.1, 19.3, 19.3, and 19.1, respectively. All data were reduced using the IPAC pipeline \cite{IPAC} with photometric calibration described in \citet{Ofek}. Several stars and asteroids were identified in the field over the four epochs, however no variable or fading candidates were identified that might be associated with FRB 140514. 

\subsection{Magellan Telescope}

Deep images of the FRB field were taken with the 6.5-m Baade telescope at Las Campanas in the $R-$ and $I-$bands on 17 May, 3 days after FRB 140514, and again on 8 July, 55 days later. These data have provided the deepest optical images of the field, with a limiting magnitude, $R,I$ = 22.5 in the first epoch and $R,I$ = 24.5 in the second epoch with a field of view of 635 arcmin$^2$. The first observation was 7 2-min exposures and the second consisted of 5 5-min exposures. These observations identified one extended object in the field that appeared with a magnitude of $R$ = 21.9 $\pm$ 0.1 in the first epoch and was not detected in the second observation through point source searches. Due to its extended nature in the observation this source has been identified as a moving object such as a satellite or debris passing through the field and was not flagged as a potentially associated candidate.

\subsection{SkyMapper}

A ToO was sent to the 1.35-m SkyMapper telescope at Siding Spring in Australia. Observations were taken on the night of the 16 May, 2 days after FRB 140514, and 23 May, 9 days after the event. The SkyMapper field of view is 5.7 deg$^2$ and both images were centred on the FRB coordinates using the H$\alpha$ filter which was in place for those nights. No variable objects were seen across the two epochs of data through difference imaging.

\subsection{Effelsberg Radio Telescope}

The field was observed at 1.4~GHz (21 cm), 2.7~GHz (11 cm), and 4.85~GHz (6 cm) using the 100-m Effelsberg Radio Telescope in Germany five days after FRB 140514. A single object was detected in the field with $S_\mathrm{1.4~GHz}$ = 447 $\pm$ 30 mJy and a spectral index of $\alpha = -0.54 \pm 0.08$. A source at this position was also visible in the NVSS with $S = 493 \pm 15$ mJy, which is consistent, indicating no change in brightness after FRB 140514. 

\subsection{Keck Spectroscopy}

Spectroscopic follow-up of XRT1, XRT2, and GMRT1 was performed on 27 May, 13 days after FRB 140514, using the Low Resolution Imaging Spectrometer (LRIS) on the 10-m Keck I telescope \cite{lris}. Based on their spectral properties, XRT1 and XRT2 were identified as AGN, and GMRT1 was identified as a starburst galaxy with a high star formation rate. The spectral features of GMRT1 were typical of a starburst galaxy and there were no strong or unexpected spectral line features that might hint at unusual activity.

\subsection{Nordic Optical Telescope Spectroscopy}

Additional spectroscopic observations were performed using the 2.5-m Nordic Optical Telescope (NOT) at La Palma for XRT1 and XRT2, 2.4 and 21.4 days after FRB 140514, respectively. Both were confirmed to be AGN. The brighter X-ray source, XRT1, is a Seyfert type 1.9 galaxy at $z$ = 0.195, and the XRT2 is an AGN with $z$ = 0.51. The spectra of both sources were not observed to evolve bewteen the two observations with NOT and Keck.

\begin{table}
\centering
\caption{Follow-up observations conducted at 12 telescopes. Limits presented are the minimum detectable magnitude or flux of each epoch. All dates are for the year 2014.}\label{tab:followup}
\tabcolsep=0.08cm
\begin{tabular}{lccc}
\hline
\hline
Telescope & Date  Start time & T+ & Limits\\
 & UTC & & \\
\hline
Parkes 	& May 14  17:14:12	& 1 s		& 1.4~GHz - 145 mJy 	\\
ATCA 	& May 15  00:10:00 	& 7 h 	 	& 5.5~GHz - 40 mJy	\\
		&					&			& 2~GHz - 60 mJy \\
Parkes 	& May 15  23:57:38 	& 6 h 52 m 	& 1.4~GHz - 145 mJy 	\\
Swift 	& May 15  01:44:43 	& 8 h 30 m 	& 8.2 $\times 10^{-15}$ erg cm$^{-2}$ s$^{-1}$ \\
GROND 	& May 15  08:49:30 	& 16 h 	& $J$ - 21.1, $H$ - 20.4,  \\
		&						&			& $K$ - 18.4 \\
Swope 	& May 15  09:57:13 	& 16 h 51 m  	& $R$ - 16	\\
iPTF 	& May 15  11:16:03 	& 18 h 11 m 	& $R$ - 19.1	\\
Swift 	& May 15  16:08:44	& 23 h 18 m 	& 3.9 $\times 10^{-15}$ erg cm$^{-2}$ s$^{-1}$ \\
GMRT 	& May 16  01:30:00 	& 1.3 d 	& 610~MHz - 125 $\upmu$Jy	\\
Effelsberg & May 16  06:50:00 & 1.4 d 	& 4.8~GHz - 2.5 mJy \\
iPTF 	& May 16  11:18:21 	& 1.7 d 	& $R$ - 19.3	\\
SkyMapper & May 16  17:57:24 & 2 d  	& H$\alpha$ - 17	\\
NOT 	& May 17  04:48:46 	& 2.4 d 	& 370 $-$ 730 nm \\
GROND 	& May 17  09:04:13 	& 2.6 d 	& $J$ - 21.1, $H$ - 20.5,	\\
		&						&			& $K$ - 18.6 \\
Swope 	& May 17  09:50:00 	& 2.6 d 	& $R$ - 16	\\
Magellan & May 17  10:11:19 & 2.6 d 	& $R$ - 22.5, $I$ - 22.5 	\\
iPTF 	& May 17  11:15:33 	& 2.7 d 	& $R$ - 19.3	\\
Effelsberg & May 18  03:50:00 	& 3.4 d 	& 2.7~GHz - 1.2 mJy \\
iPTF 	& May 19  11:23:52 	& 4.7 d 	& $R$ - 19.1	\\
Effelsberg & May 21  05:35:00 & 7.5 d 	& 1.4~GHz - 1.2 mJy \\
SkyMapper & May 23  17:45:48 	& 9 d	& H$\alpha$ - 17	\\
Keck 	& May 27  14:06:22 	& 12.8 d 	& 30 $-$ 1000 nm	\\
Swift 	& June 02  00:06:02 	& 18.3 d 	& 6.35 $\times 10^{-15}$ erg cm$^{-2}$ s$^{-1}$ \\
GMRT 	& June 03  00:20:00 	& 19.3 d 	& 1390~MHz - 61 $\upmu$Jy	\\
NOT 	& June 05  03:51:09 	& 21.4 d	& 370 $-$ 730 nm	\\
GMRT 	& June 08  20:30:00 	& 24.1 d	& 610 MHz - 150 $\upmu$Jy 	\\
Parkes 	& June 24  14:36:40 	& 41 d 		& 1.4~GHz - 145 mJy \\
Magellan 	& July 8  07:34:44 	& 55 d 	& $R$ - 24.5, $I$ - 24.5	\\
Parkes 	& July 27  12:14:00 	& 74 d 	& 145 mJy \\
\hline
\end{tabular}

\end{table}

\section{Interpretation and Discussion}\label{sec:discussion}

Over the various epochs and wavelengths detailed in Section~\ref{sec:followup}, no afterglow-like variable counterparts were detected that could be identified as a candidate host or progenitor associated with the radio observation of FRB 140514. Here we consider the possible sources and mechanisms that might produce the observed behavior and set limits on FRB detectability at other wavelengths on hour-to-day timescales.

\subsection{Polarization}\label{sec:polarization}

Any progenitor theory of FRBs must explain the observed polarization; several possibilities exist for this FRB. We consider three here:
\begin{itemize}\itemsep0pt
\item \textit{Case 1:} The emission is intrinsically only circularly polarized, as observed, perhaps also with low linear polarization.
\item \textit{Case 2:} The emission is intrinsically linearly and circularly polarized, but the linear polarization was undetectable due to bandwidth depolarization by severe Faraday rotation.
\item \textit{Case 3:} The emission is intrinsically unpolarized and circular polarization is scintillation-induced.
\end{itemize}

\textit{Case 1.} Few sources observed at radio frequencies produce only circularly polarized emission. The flare star AD Leonis has been observed to produce 90-100\% circularly polarized radio emission coincident with optical flares \cite{Osten1,Osten2}; the brown dwarf TVLM 513$-$46546 and the Sun have also been observed to emit 100\% circularly polarized bursts at GHz frequencies, both attributed to cyclotron maser emission \cite{Hallinan,SolarBurst}. These radio bursts typically last seconds to minutes with no frequency-dependent time delays comparable to the $\nu^{-2}$ dispersive sweep seen for FRBs. The level of circular polarization (CP) in FRB 140514 ($\sim21\%$) is also much lower than that observed in other cases. Some AGN have been observed with more circular polarization than linear \cite{Homan}, however the overall levels of CP were much lower (typically $\geq 0.3\%$) and would not have been detected here. Some single pulses from pulsars have been observed with high fractional circular polarisation and a small linear component and this FRB may represent such a state \cite{Stefan0437,LevinMagnetar}.

\textit{Case 2.} The maximum RM for the search in $\S$\ref{sec:radio} describes the complete depolarization of a 100\% linearly polarized source. A weaker level of linear polarization might have been depolarized at these frequencies by RMs of order $\sim10^{4}$ rad m$^{-2}$ and greater, however such values are still several orders of magnitude greater than theorized for Faraday rotation in the IGM \cite{Akahori}. Additionally, if the source originated at some redshift and the emission observed at 1.4~GHz was redshifted into our observing band, the Faraday rotation at the source would need to be much higher given the frequency of emission. The electron density component of the RM is constrained by the source DM, thus high magnetic fields are required to produce the necessary rotation of the plane of linear polarisation. Such high rotation measures are incredibly rare, but have been observed in the magnetar PSR J1745-2900 near the Galactic Centre \cite{GCmagnetar,ShannonGC}. The required path-averaged magnetic field to produce these rotation measures would be $\geq$ 250 $\upmu$G for FRB 140514. A source located within 1 pc of the center of its host galaxy, with the host contributing 100 pc cm$^{-3}$ to the total DM, could produce path-averaged magnetic field strengths of 10-100 $\upmu$G. FRB 140514 would need to have originated in a region of similar magnetic field strength to produce the necessary Faraday rotation. It is also worth noting that observations of a radio-loud magnetar have shown the occurence of an infrequent state in which the emission is highly circularly polarized with a lower-than-average linear component \cite{LevinMagnetar}. With a sufficiently strong integrated magnetic field along the line of sight, the linear component could be completely depolarized for a magnetar flare. Such a flare would still be in good agreement with the models and conditions put forward in \citet{Kulkarni14}.

\textit{Case 3.} It has been theorized that scintillation-induced CP may arise in an intrinsically unpolarized source \cite{MacquartCP}. In the diffractive regime - for pulsars and other Galactic sources - CP up to $\sim15\%$ may be induced by a birefringent medium at low frequencies, and in the refractive regime $\sim0.1\%$ may be induced for extragalactic sources observed at higher radio frequencies. Such scintillation requires the presence of an RM gradient across the turbulent region, which we cannot constrain with available data. The $\sim15\%$ CP in the \citet{MacquartCP} simulation was derived from the Vela pulsar, an extreme example within the pulsar population, being bright, young, and surrounded by a turbulent and high-velocity medium \cite{VelaDM,PetroffDM}. Such an object would very likely be detected in future follow-up of the detection position.

Of the three posibilities presented here, \textit{Case 3} requires very specialized Galactic conditions to produce CP close to the level observed in this work and is not the best model to describe the observed FRB circular polarization. Both \textit{Case 1} and \textit{Case 2} require CP intrinsic to the source, and vary only in on the predicted level of linear polarization.

Recent theoretical work has speculated that mechanisms to produce sufficiently high brightness temperatures for observed FRBs require beaming of coherent emission \cite{Katz2014} which would produce intrinsic linear polarization, making \textit{Case 2} more appealing. The required brightness temperature $T_\mathrm{B}$ for FRB 140514, ignoring relativistic effects, can be calculated using

\begin{equation}\label{eq:tb}
T_\mathrm{B} \simeq 10^{36} \mathrm{K} \left(\frac{S_\mathrm{peak}}{\mathrm{Jy}}\right) \left(\frac{\mathrm{GHz}}{\nu}\right)^{2} \left(\frac{\mathrm{ms}}{\Delta t}\right)^{2} \left(\frac{d}{\mathrm{Gpc}}\right)^{2} \frac{(1+z)^4}{\gamma^2}
\end{equation}

\noindent where $S_\mathrm{peak}$ is the flux density, $\nu$ is the centre observing frequency, $\Delta t$ is the pulse width, and $d$ is the co-moving distance \cite{PulsarHandbook}. For FRB 140514 we estimate a brightness temperature $T_\mathrm{B} = 5.3 \times 10^{35}$ K, not including relativistic effects. Such a high brightness temperature is beyond the regime of pulses from typical pulsars but approaches what is seen in the brightest nanoshot pulses from the Crab pulsar (10$^{41}$ K) \cite{Hankins}. Temperatures in this regime preclude synchrotron emission, thus making it likely that the pulse emission is coherent \cite{Readhead1994}. However, no linear polarization was detected, an unexpected result given previous observations of coherent emission mechanisms at such high brightness temperatures \cite{CrabPol}.

An additional theory of FRB origins put forward by \citet{Mottez2014} suggests that FRBs are created in the Alfv\'en wings of a planet orbiting a neutron star within the pulsar wind via the electron cyclotron maser (ECM) instability at Gpc distances. This theory predicts stong circular polarization such as the 100\% circular emission seen in other objects that emit via ECM such as the M dwarves observed by \citet{Hallinan2008}. While this is the only current theory that explicitly predicts circularly polarised emission from FRBs, a significant fraction of the intrinsic circular polarisation ($>$50\%) would have been lost by some unknown mechanism for FRB 140514 to explain the observed polarised profile.

We then conclude that \textit{Case 1} or \textit{Case 2} may be the best explanation of the observed polarization for FRB 140514, although, the non-detection of linear polarization at 1.4~GHz might require extremely high magnetic fields to produce the necessary Faraday rotation, possibly near a galactic center. 

In the future, more sensitive measurements of FRB polarization may be possible with coherent baseband capture buffers such as the CASPER Parkes Swinburne Recorder (CASPSR) installed on the center beam of the multibeam receiver at Parkes, and those being designed for the Square Kilometre Array (SKA).

\subsection{Possible connection with FRB 110220}\label{sec: FRB 110220}

FRB 140514 was discovered in radio follow-up observations of a previous FRB event, FRB 110220, published in \citet{Thornton13}. FRB 110220 was the most extensively analyzed FRB in this sample as it was the brightest, detected with a S/N of 49, an extremely high DM of 944 cm$^{-3}$ pc, a $\nu^{-2.003\pm0.006}$ dispersion relation, and a significant scattering tail.

FRB 140514 was discovered in a grid pointing around the position of  FRB 110220: the centre beam of the receiver was centered 9$'$ away from the detection beam position for FRB 110220. Given the overlap of the 14$'$ beam on-sky between the two FRBs it is tempting to make an association. A few considerations must be made before attributing both events to the same source - the probability of detecting a new source given the FRB rate and the time on-sky for these observations, the physical mechanism necessary to produce a large change in dispersion measure over the time between detections, and the time of day and position of the telescope at the time of each detection.

The probability of detecting a new FRB in our observations can be calculated using the formula 
\begin{equation}\label{eq:prob}
P(\mathrm{N}|\mathrm{M}) = \alpha^{\mathrm{N}} (1+\alpha)^{-(1+\mathrm{M} + \mathrm{N})} \frac{(\mathrm{M}+\mathrm{N})!}{\mathrm{M}! \: \mathrm{N}!},
\end{equation}
derived in \citet{Petroff14} to find the likelihood of detecting N FRBs in our survey based on M detections in a previous survey with a ratio between their cumulative time on sky of $\alpha$. This FRB campaign (with extra time granted by the scheduler) has spent 85 hours on-sky over the duration of the survey. The probability of finding a new FRB in these data, given the FRB rate from \citet{Thornton13} is $\sim$32\%, if FRBs are non-repeating. 

Recent results from \citet{SarahFRB} and \citet{Petroff14} have both addressed the possibility of a lower rate based on results from searches at lower Galactic latitudes, possibly due to foreground effects. Using the rate derived in \citet{SarahFRB} we estimate a much lower detection probability of $\sim$ 2\% from our survey. However, \citet{SarahFRB} note that the number of detectable FRBs may be heavily latitude dependent, with different rates applying for surveys at high and low Galactic latitude. In the absence of an expression for this dependence we will continue to use the probability derived from the results of \citet{Thornton13} as their survey explicity sampled the region around FRB 140514. Based on comparison with previous rate estimates it is therefore not unexpected that we would find a new FRB in this project.

FRB 110220 was observed with a DM of 944.7 cm$^{-3}$ pc, compared to a DM of 562 cm$^{-3}$ pc for FRB 140514, a difference of 380 cm$^{-3}$ pc for events separated by 1179 days. This is a temporal variation in the DM which is four orders of magnitude greater than what is seen for pulsars in the most turbulent environments (e.g. $|$dDM/d$t$$|$ = 0.18$\pm$1 cm$^{-3}$ pc yr$^{-1}$ for PSR J1833-0827, \citet{PetroffDM}).

In order for such large DM variations to be observed for these two FRBs, most of the dispersive medium for  FRB 110220 would be local to the source and thus extremely dense. \citet{Dennison14} and \citet{Artem14} have both argued that if the majority of the DM were produced in a dense plasma around the emission region, such as in a stellar corona, the observed dispersion relation would be poorly fit by a $\nu^{-2}$ relation, which would have been easily detected in the analysis conducted by \citet{Thornton13} for FRB 110220 and the analysis conducted here for FRB 140514. Additionally, FRB 140514 was discovered at 03:14 AEST local time at a telescope orientation in azimuth and elevation of (84.9$^\circ$, 30.3$^\circ$) while FRB 110220 occurred at 11:52 AEST local time at a telescope orientation of (29.9$^\circ$, 66.3$^\circ$).

We conclude that FRB 110220 and FRB 140514 are different sources, and their proximity is purely due to sampling bias in our choice of observing location. This proximity does not affect the proposed cosmological origin of FRBs.

\subsection{Limits on a varying counterpart}\label{sec:afterglow}

In explosions such as supernovae (SNe) and superluminous supernovae (SLSNe), or in long gamma-ray bursts (GRBs), a counterpart is detectable as an object of varying brightness in subsequent observations. Variations in brightness would be observable on timescales of hours (for a long GRB), days (for typical SNe) or weeks (for a SLSN). Such variations were not seen in our data in X-ray, near-infrared, optical, or radio regimes. Additionally, no gamma-ray emission from the field was observed by the interplanetary network (IPN) in either the hard or soft gamma-ray energy bands, ruling out all but soft, short gamma-ray bursts (V. Pal'shin, priv. comm.). Therefore, no variable counterpart or related transient emission was observed in association with FRB 140514. From the extensive dataset collected in this analysis we can then place limits on the magnitude of any potential afterglow for FRB 140514. 

\begin{figure*}
\includegraphics[width=18cm]{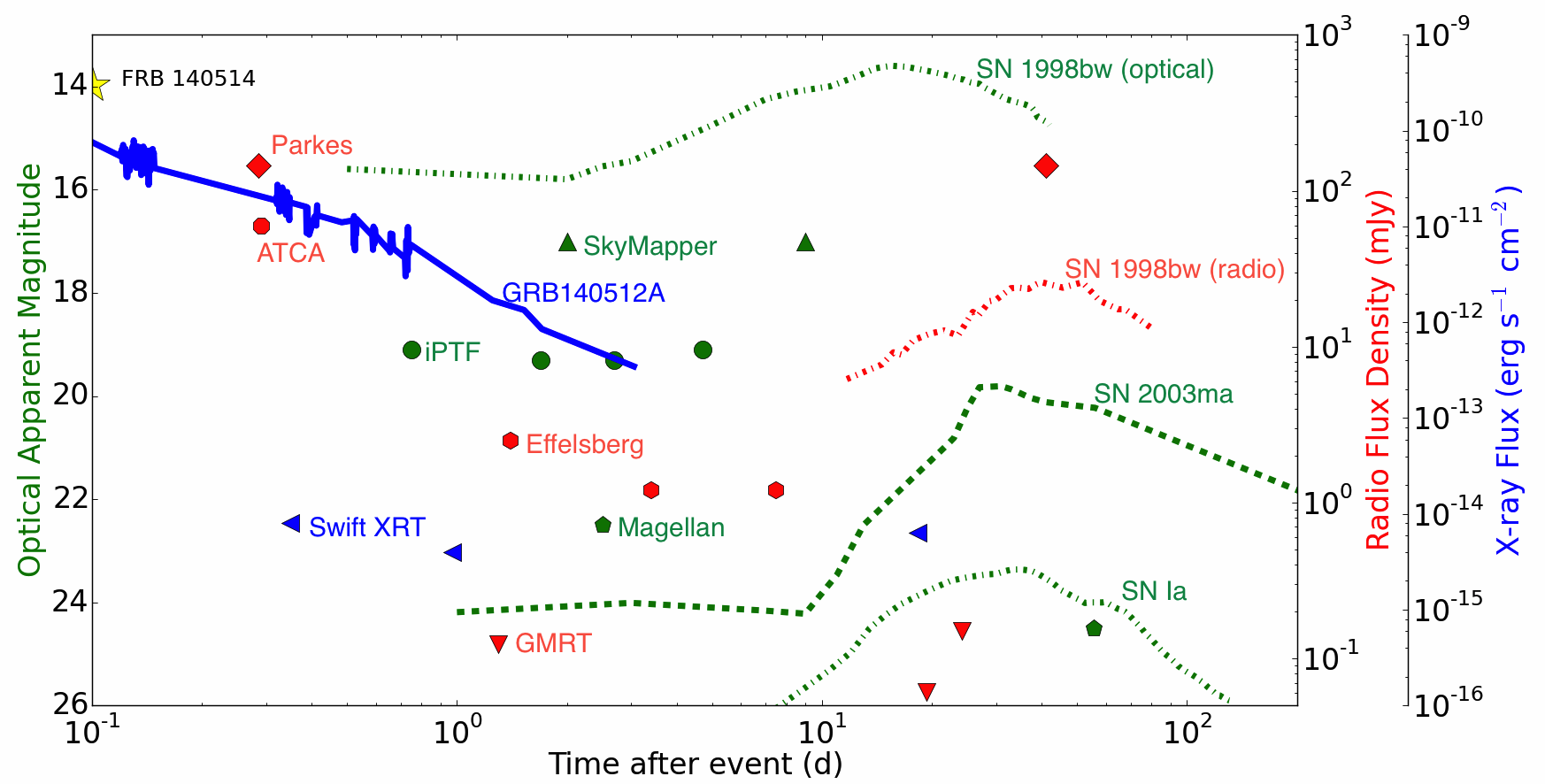}
\caption{The limits for optical in apparent magnitude (green), radio flux density in mJy (red), and X-ray flux in erg cm$^{-2}$ s$^{-1}$ (blue) of our observations of the field of FRB 140514 from 8 telescopes that fully sampled the Parkes beam. Colors of data points refer to the axis scale of the same color. Light curves from GRB140512A ($z$ = 0.725), 1.4 GHz radio data and $R-$band optical data for supernova SN1998bw ($z \sim$0.008), $R-$band data for superluminous supernova SN2003ma ($z$ = 0.289) and an $R-$band light curve for a typical type-Ia SN ($z$ = 0.5) have been included for reference \cite{SwiftOnline,SLSN,Kulkarni14,Galama}. \label{fig:lightcurves}}
\end{figure*}

We place the limits on related transient emission at 20.0 in $J$-band (1.65 $\upmu$m), 19.2 in $H$-band (1.2 $\upmu$m), 18.6 in $K$-band (2.2 $\upmu$m), 24.5 in $R$-band, 24.5 in $I$-band, 1.5 mJy at 4.8 GHz, 60 mJy at 1-3 GHz, and 125 $\upmu$Jy at 610 MHz. We compare these limits to light curves of known variable supernovae and gamma-ray bursts (Figure~\ref{fig:lightcurves}) and find that many nearby sources would have been detected in this analysis \cite{SwiftOnline,SLSN,RadioSN,Galama}. We rule out local superluminous supernovae and nearby ($z <$ 0.3) type Ia supernovae, as well as slow transients with variations greater than 2 mag AB between our epochs of observation. We also rule out an FRB association with long GRBs.

Such constraints make associations between FRB 140514 and some SNe or long GRBs highly unlikely. Any theoretical description of FRB emission must adhere to these constraints. 

\subsection{Host galaxies}\label{sec:hosts}
%% Suggested by SBS
Given the large error radius of the Parkes beam (7.2$'$) an identification of a host galaxy in the original radio detection was not possible and no other association was detected at other wavelengths as no variable sources were identified in the field. In our follow-up observations three sources of interest were detected in the field of FRB 140514 by \textit{Swift} and the GMRT: two X-ray luminous AGN and one radio-loud starburst galaxy. The identification of AGN within a 15$'$ beam is not unexpected; approximately 3 AGN are expected to occur within the area of the Parkes beam out to $z = 2$ based on cosmological distribution alone \cite{AGNdensity}. Finding a starburst galaxy in our field is also likely based on galaxy distribution studies \cite{Hirashita}. Therefore, we are unable to make any robust connection between the FRB and any other objects identified in the field on probability arguments alone. 

\section{Conclusions}\label{sec:conclusion}
We report here on an FRB discovered in real-time at 1.4~GHz at the Parkes radio telescope. FRB 140514 was the first real-time FRB detection and the first with polarization information. The pulse was observed to be 21$\pm$7\% (3-$\sigma$) circularly polarized with a 1-$\sigma$ upper limit of 10\% on linear polarization which was not detected. We coordinated the fastest and largest follow-up effort ever undertaken for an FRB, with data from 12 telescopes. No associated slow transient, progenitor, or host galaxy was identified, effectively ruling out any association between this FRB and a supernova at $z<0.3$ or long gamma-ray burst. A tighter constraint on FRB origins in the future will require not only robust and immediate triggering or commensal observing at multiple observatories, but also improved sky localisation of radio pulses within FRB and pulsar surveys. We note that any theory of FRB origin must satisfy the polarization, brightness temperature, and afterglow limits put forward in this analysis.

\section*{Public Data Release}

The data presented in this paper are made available through Research Data Australia\footnote{https://researchdata.ands.org.au/fast-radio-burst-frb-140514/468269} and can be processed using the publicly available \textsc{Heimdall} single pulse processing software and the \textsc{psrchive} software package.

\section*{Acknowledgements} 

The Parkes radio telescope and the Australia Telescope Compact Array are part of the Australia Telescope National Facility which is funded by the Commonwealth of Australia for operation as a National Facility managed by CSIRO. Parts of this research were conducted by the Australian Research Council Centre of Excellence for All-sky Astrophysics (CAASTRO), through project number CE110001020. We thank the staff of the GMRT that made these observations possible. GMRT is run by the National Centre for Radio Astrophysics of the Tata Institue of Fundamental Research. Research with the ANU SkyMapper telescope is supported in part through ARC Discovery Grant DP120101237. We thank the Carnegie Supernova Project team (PI M. Phillips) and intermediate Palomar Transient Factory team (PI S. Kulkarni) for promptly taking follow-up data. Part of the funding for GROND (both hardware as well as personnel) was generously granted from the Leibniz-Prize to Prof. G. Hasinger (DFG grand HA 1850/28-1). The Dark Cosmology Centre is supported by the Danish National Research council. We thank A. Krauss for prompt observations with the Effelsberg radio telescope. Partly based on observations made with the Nordic Optical Telescope, operated by the Nordic Optical Telescope Scientific Association at the Observatorio del Roque de los Muchachos, La Palma, Spain, of the Instituto de Astrof\'isica de Canarias.

We thank the anonymous referee for valuable input which improved the clarity of this paper. EP would like to thank M. Murphy, J. Cooke, and C. Vale for useful discussion and valuable comments. NDRB is supported by a Curtin Research Fellowship. CD acknowledges support through EXTraS, funded from the European Union's Seventh Framework Programme for research, technological development and demonstration under grant agreement no 607452. MMK acknowledges generous support from the Hubble Fellowship and Carnegie-Princeton Fellowship. DM acknowledged the Instrument Center for Danish Astrophysics (IDA) for support. EOO is incumbent of the Arye Dissentshik career development chair and is grateful to support by grants from the Willner Family Leadership Institute Ilan Gluzman (Secaucus NJ), Israeli Ministry of Science, Israel Science Foundation, Minerva, Weizmann-UK and the I-CORE Program of the Planning and Budgeting Committee and The Israel Science Foundation. Support for DAP was provided by NASA through Hubble Fellowship grant HST-HF-51296.01-A awarded by the Space Telescope Science Institute, which is operated by the Association of Universities for Research in Astronomy, Inc., for NASA, under contract NAS 5-26555. BPS, CW, and PT, acknowledge funding from the ARC via CAASTRO and grand LF0992131.

\bibliographystyle{mnras}
\bibliography{journals,FRB140514}

\begin{thebibliography}{}

\bibitem[\protect\citeauthoryear{{Akahori} \& {Ryu}}{{Akahori} \&
  {Ryu}}{2010}]{Akahori}
{Akahori} T.,  {Ryu} D., 2010, ApJ, 723, 476

\bibitem[\protect\citeauthoryear{{Burke-Spolaor} \&
  {Bannister}}{{Burke-Spolaor} \& {Bannister}}{2014}]{SarahFRB}
{Burke-Spolaor} S.,  {Bannister} K.~W., 2014, \apj, 792, 19

\bibitem[\protect\citeauthoryear{{Condon} et~al.}{{Condon} et~al.}{1998}]{NVSS}
{Condon} J.~J., {Cotton} W.~D., {Greisen} E.~W., {Yin} Q.~F., {Perley} R.~A.,
  {Taylor} G.~B.,  {Broderick} J.~J., 1998, \aj, 115, 1693

\bibitem[\protect\citeauthoryear{{Cordes} \& {Lazio}}{{Cordes} \&
  {Lazio}}{2002}]{Cordes02}
{Cordes} J.~M.,  {Lazio} T.~J.~W., 2002, ArXiv Astrophysics e-prints
  (arXiv:0207156)

\bibitem[\protect\citeauthoryear{{Deng} \& {Zhang}}{{Deng} \&
  {Zhang}}{2014}]{Deng2014}
{Deng} W.,  {Zhang} B., 2014, \apjl, 783, L35

\bibitem[\protect\citeauthoryear{{Dennison}}{{Dennison}}{2014}]{Dennison14}
{Dennison} B., 2014, \mnras, 443, L11

\bibitem[\protect\citeauthoryear{{Eatough} et~al.}{{Eatough}
  et~al.}{2013}]{GCmagnetar}
{Eatough} R.~P. et~al., 2013, Nature, 501, 391

\bibitem[\protect\citeauthoryear{{Evans} et~al.}{{Evans}
  et~al.}{2007}]{SwiftOnline}
{Evans} P.~A. et~al., 2007, \aap, 469, 379

\bibitem[\protect\citeauthoryear{{Fiore} et~al.}{{Fiore}
  et~al.}{2003}]{AGNdensity}
{Fiore} F. et~al., 2003, \aap, 409, 79

\bibitem[\protect\citeauthoryear{{Galama} et~al.}{{Galama}
  et~al.}{1998}]{Galama}
{Galama} T.~J. et~al., 1998, \nat, 395, 670

\bibitem[\protect\citeauthoryear{{Gao}, {Li}, \& {Zhang}}{{Gao}
  et~al.}{2014}]{Gao14}
{Gao} H., {Li} Z.,  {Zhang} B., 2014, \apj, 788, 189

\bibitem[\protect\citeauthoryear{{Gehrels} et~al.}{{Gehrels}
  et~al.}{2004}]{swift}
{Gehrels} N. et~al., 2004, \apj, 611, 1005

\bibitem[\protect\citeauthoryear{{Greiner} et~al.}{{Greiner}
  et~al.}{2008}]{grond}
{Greiner} J. et~al., 2008, \pasp, 120, 405

\bibitem[\protect\citeauthoryear{{Hallinan} et~al.}{{Hallinan}
  et~al.}{2008}]{Hallinan2008}
{Hallinan} G., {Antonova} A., {Doyle} J.~G., {Bourke} S., {Lane} C.,  {Golden}
  A., 2008, \apj, 684, 644

\bibitem[\protect\citeauthoryear{{Hallinan} et~al.}{{Hallinan}
  et~al.}{2007}]{Hallinan}
{Hallinan} G. et~al., 2007, \apjl, 663, L25

\bibitem[\protect\citeauthoryear{{Hamilton}, {Hall}, \& {Costa}}{{Hamilton}
  et~al.}{1985}]{VelaDM}
{Hamilton} P.~A., {Hall} P.~J.,  {Costa} M.~E., 1985, MNRAS, 214, 5P

\bibitem[\protect\citeauthoryear{{Hankins} \& {Eilek}}{{Hankins} \&
  {Eilek}}{2007}]{Hankins}
{Hankins} T.~H.,  {Eilek} J.~A., 2007, \apj, 670, 693

\bibitem[\protect\citeauthoryear{{Hankins} et~al.}{{Hankins}
  et~al.}{2003}]{CrabPol}
{Hankins} T.~H., {Kern} J.~S., {Weatherall} J.~C.,  {Eilek} J.~A., 2003, \nat,
  422, 141

\bibitem[\protect\citeauthoryear{{Hirashita} et~al.}{{Hirashita}
  et~al.}{1999}]{Hirashita}
{Hirashita} H., {Takeuchi} T.~T., {Shibai} H.,  {Ohta} K., 1999, PASJ, 51, 81

\bibitem[\protect\citeauthoryear{{Homan} \& {Lister}}{{Homan} \&
  {Lister}}{2006}]{Homan}
{Homan} D.~C.,  {Lister} M.~L., 2006, ApJ, 131, 1262

\bibitem[\protect\citeauthoryear{{Hotan}, {van Straten}, \&
  {Manchester}}{{Hotan} et~al.}{2004}]{Hotan2004}
{Hotan} A.~W., {van Straten} W.,  {Manchester} R.~N., 2004, \pasa, 21, 302

\bibitem[\protect\citeauthoryear{{Ioka}}{{Ioka}}{2003}]{Ioka03}
{Ioka} K., 2003, ApJ Letters, 598, L79

\bibitem[\protect\citeauthoryear{{Katz}}{{Katz}}{2014}]{Katz2014}
{Katz} J.~I., 2014, Phys Rev D, 89, 103009

\bibitem[\protect\citeauthoryear{{Keith} et~al.}{{Keith}
  et~al.}{2010}]{Keith10}
{Keith} M.~J. et~al., 2010, MNRAS, 409, 619

\bibitem[\protect\citeauthoryear{{Kulkarni} et~al.}{{Kulkarni}
  et~al.}{1998}]{RadioSN}
{Kulkarni} S.~R. et~al., 1998, \nat, 395, 663

\bibitem[\protect\citeauthoryear{{Kulkarni} et~al.}{{Kulkarni}
  et~al.}{2014}]{Kulkarni14}
{Kulkarni} S.~R., {Ofek} E.~O., {Neill} J.~D., {Zheng} Z.,  {Juric} M., 2014,
  ArXiv e-prints:1402.4766

\bibitem[\protect\citeauthoryear{{Laher} et~al.}{{Laher} et~al.}{2014}]{IPAC}
{Laher} R.~R. et~al., 2014, \pasp, 126, 674

\bibitem[\protect\citeauthoryear{{Law} et~al.}{{Law} et~al.}{2009}]{ptf}
{Law} N.~M. et~al., 2009, \pasp, 121, 1395

\bibitem[\protect\citeauthoryear{{Levin} et~al.}{{Levin}
  et~al.}{2012}]{LevinMagnetar}
{Levin} L. et~al., 2012, \mnras, 422, 2489

\bibitem[\protect\citeauthoryear{{Loeb}, {Shvartzvald}, \& {Maoz}}{{Loeb}
  et~al.}{2014}]{Loeb14}
{Loeb} A., {Shvartzvald} Y.,  {Maoz} D., 2014, MNRAS, 439, L46

\bibitem[\protect\citeauthoryear{{Lorimer} et~al.}{{Lorimer}
  et~al.}{2007}]{Lorimer07}
{Lorimer} D.~R., {Bailes} M., {McLaughlin} M.~A., {Narkevic} D.~J.,  {Crawford}
  F., 2007, Science, 318, 777

\bibitem[\protect\citeauthoryear{{Lorimer} \& {Kramer}}{{Lorimer} \&
  {Kramer}}{2004}]{PulsarHandbook}
{Lorimer} D.~R.,  {Kramer} M., 2004, {Handbook of Pulsar Astronomy}, Vol.~4.
\newblock Cambridge University Press

\bibitem[\protect\citeauthoryear{{Lyubarsky}}{{Lyubarsky}}{2014}]{Lyubarsky2014}
{Lyubarsky} Y., 2014, \mnras, 442, L9

\bibitem[\protect\citeauthoryear{{Macquart} \& {Melrose}}{{Macquart} \&
  {Melrose}}{2000}]{MacquartCP}
{Macquart} J.-P.,  {Melrose} D.~B., 2000, \apj, 545, 798

\bibitem[\protect\citeauthoryear{{Manchester} et~al.}{{Manchester}
  et~al.}{2005}]{psrcat}
{Manchester} R.~N., {Hobbs} G.~B., {Teoh} A.,  {Hobbs} M., 2005, AJ, 129, 1993

\bibitem[\protect\citeauthoryear{{Melrose} \& {Dulk}}{{Melrose} \&
  {Dulk}}{1982}]{SolarBurst}
{Melrose} D.~B.,  {Dulk} G.~A., 1982, \apj, 259, 844

\bibitem[\protect\citeauthoryear{{Monet} et~al.}{{Monet}
  et~al.}{2003}]{USNOcatalog}
{Monet} D.~G. et~al., 2003, \aj, 125, 984

\bibitem[\protect\citeauthoryear{{Morris} et~al.}{{Morris}
  et~al.}{2002}]{gridding}
{Morris} D.~J. et~al., 2002, \mnras, 335, 275

\bibitem[\protect\citeauthoryear{{Mottez} \& {Zarka}}{{Mottez} \&
  {Zarka}}{2014}]{Mottez2014}
{Mottez} F.,  {Zarka} P., 2014, \aap, 569, A86

\bibitem[\protect\citeauthoryear{{Ofek} et~al.}{{Ofek} et~al.}{2012}]{Ofek}
{Ofek} E.~O. et~al., 2012, \pasp, 124, 62

\bibitem[\protect\citeauthoryear{{Oke} et~al.}{{Oke} et~al.}{1995}]{lris}
{Oke} J.~B. et~al., 1995, \pasp, 107, 375

\bibitem[\protect\citeauthoryear{{Os{\l}owski} et~al.}{{Os{\l}owski}
  et~al.}{2014}]{Stefan0437}
{Os{\l}owski} S., {van Straten} W., {Bailes} M., {Jameson} A.,  {Hobbs} G.,
  2014, \mnras, 441, 3148

\bibitem[\protect\citeauthoryear{{Osten} \& {Bastian}}{{Osten} \&
  {Bastian}}{2006}]{Osten1}
{Osten} R.~A.,  {Bastian} T.~S., 2006, ApJ, 637, 1016

\bibitem[\protect\citeauthoryear{{Osten} \& {Bastian}}{{Osten} \&
  {Bastian}}{2008}]{Osten2}
{Osten} R.~A.,  {Bastian} T.~S., 2008, ApJ, 674, 1078

\bibitem[\protect\citeauthoryear{{Petroff} et~al.}{{Petroff}
  et~al.}{2013}]{PetroffDM}
{Petroff} E., {Keith} M.~J., {Johnston} S., {van Straten} W.,  {Shannon} R.~M.,
  2013, \mnras, 435, 1610

\bibitem[\protect\citeauthoryear{{Petroff} et~al.}{{Petroff}
  et~al.}{2014}]{Petroff14}
{Petroff} E. et~al., 2014, \apjl, 789, L26

\bibitem[\protect\citeauthoryear{{Rau} et~al.}{{Rau} et~al.}{2009}]{Rau09}
{Rau} A. et~al., 2009, \pasp, 121, 1334

\bibitem[\protect\citeauthoryear{{Readhead}}{{Readhead}}{1994}]{Readhead1994}
{Readhead} A.~C.~S., 1994, \apj, 426, 51

\bibitem[\protect\citeauthoryear{{Rest} et~al.}{{Rest} et~al.}{2011}]{SLSN}
{Rest} A. et~al., 2011, \apj, 729, 88

\bibitem[\protect\citeauthoryear{{Shannon} \& {Johnston}}{{Shannon} \&
  {Johnston}}{2013}]{ShannonGC}
{Shannon} R.~M.,  {Johnston} S., 2013, \mnras, 435, L29

\bibitem[\protect\citeauthoryear{{Spitler} et~al.}{{Spitler}
  et~al.}{2014}]{Spitler14}
{Spitler} L.~G. et~al., 2014, \apj, 790, 101

\bibitem[\protect\citeauthoryear{{Staveley-Smith} et~al.}{{Staveley-Smith}
  et~al.}{1996}]{multibeam}
{Staveley-Smith} L. et~al., 1996, PASA, 13, 243

\bibitem[\protect\citeauthoryear{{Thornton} et~al.}{{Thornton}
  et~al.}{2013}]{Thornton13}
{Thornton} D. et~al., 2013, Science, 341, 53

\bibitem[\protect\citeauthoryear{{Tuntsov}}{{Tuntsov}}{2014}]{Artem14}
{Tuntsov} A.~V., 2014, MNRAS, 441, L26

\bibitem[\protect\citeauthoryear{{Wright}}{{Wright}}{2006}]{CosmologyCalc}
{Wright} E.~L., 2006, \pasp, 118, 1711

\end{thebibliography}

\end{document}